\documentclass{IEEEcsmag}

\usepackage{hyperref}

\usepackage{upmath}
\usepackage[dvipsnames,table,xcdraw]{xcolor}
\usepackage{color}
\usepackage[most]{tcolorbox}

\newtcolorbox{guidelinebox}[2][]{enhanced,
    sharp corners,
    colback=white,
    colbacktitle=white,
    coltitle=black,
    boxrule=1pt,
    left=5mm,
    right=5mm,
    bottom=2mm,
    top=4mm,
    boxed title style={colframe=white},
    attach boxed title to top center={yshift=-3mm},
    center title,
    title=#2,#1
}

\jvol{--}
\jnum{--}
\paper{--}
\jmonth{--}
\jname{IEEE Software}
\pubyear{2022}

\begin{document}

\sptitle{--}
\editor{--}

\title{The Present and Future of Bots in Software Engineering}

\author{Emad Shihab}
\affil{Concordia University, Canada}

\author{Stefan Wagner}
\affil{University of Stuttgart, Germany}

\author{Marco A. Gerosa}
\affil{Northern Arizona University, USA}

\author{Mairieli Wessel}
\affil{Radboud University, The Netherlands}

\author{Jordi Cabot}
\affil{ICREA, Spain}

%\markboth{Special Issue on Bots in Software Engineering}

% There is no abstract as far as I can see in the examples
%\begin{abstract}

%\end{abstract}

\maketitle

\chapterinitial{Software engineering bots} are applications able to react to external stimuli such as events triggered by tools or messages posted by users and run automated tasks in response, working as an interface between users and services. Bots often include conversational capabilities to interact with end-users through textual messages (in chatbots) or speech (in voicebots) in the same communication channels as their human counterparts. Bots can support technical and social activities in software engineering, including communication and decision-making.

We are witnessing a massive adoption of bots in a variety of domains, including e-commerce, customer service, and education. Software development is no exception \cite{StoreyZ16, ErlenhovN020}. Given the essential complexity of software projects and the large community of people around them (stakeholders, designers, developers and, let's not forget, end-users), there are plenty of opportunities for bots to jump in and tame this complexity by (semi)automating repetitive tasks. We often see bots working on software repositories, e.g., to manage pull requests, acting as Q\&A bots, e.g., for information retrieval, or integrated in software development environments, e.g., automating bug repair \cite{santhanam2022}.

Automation is even more relevant for open-source projects, which typically face sustainability issues. The adoption of bots may help free some responsibilities of the open-source maintainers and allow them to focus on the most critical tasks, benefiting the long-term health of open source. In open-source (or inner-source) projects, bots can leverage the public availability of software assets, including source code, discussions, issues and comments, to target more significant contributions.

This special issue offers a perspective on the current role of bots in software engineering.

\section{Overview of papers in this special issue}
% each of us summarizes their papers

%The papers in this special issue provide a comprehensive perspective of the state of the art of the use and benefits of bots in Software Engineering.

Zimmerman et al., in their paper ``Extending the team with a project-specific bot,'' report their experience developing and maintaining a custom bot named Coq bot, which was built to support the Coq team (circa 40 developers and hundreds of contributors). The bot was initially developed to automate the synchronization between pull requests opened on a GitHub repository and branches on a GitLab mirror. Based on user feedback, the bot evolved to execute other tasks, including merging a pull request, keeping track of pull requests with merge conflicts, and backporting pull requests. The authors note that relying on familiar technology and straightforward and extensible architecture choices can ease the maintenance of a bot by facilitating the onboarding of new bot maintainers.

The adoption and characterization of bots in open-source projects is the topic of the article ``From Specialized Mechanics to Project Butlers: the Usage of Bots in Open Source Development'' by Wang and Redmiles. In this article, the authors sampled the top 1000 most popular (using the number of stars as a popularity metric) software development repositories on GitHub and studied whether the project employed software bots and, if so, what types of tasks those bots were helping with. As part of their conclusions, they highlight that over 60\% of open-source projects do use bots, even though these bots often focus on automating simple tasks. The authors note that these bots are typically rule-based reacting to certain events they have subscribed to and show very limited interactive capabilities. 

Cogo and Hassan focus in their article ``Understanding the Customization of Dependency Bots: The Case of Dependabot'' on a popular bot used on GitHub named \emph{Dependabot}. This bot is used for checking and updating dependencies to libraries used in a project. The authors analyzed almost 500 projects that use Dependabot and have corresponding configuration files. The authors conclude that customizing a bot's behavior can help in reducing noise but might also limit the bot's usefulness as certain features might stop working. Therefore, designers of bots should be careful in considering the trade-off between allowing bot users to configure the bot and the interaction with features that the bot offers. In general, configurations should be as simple as possible since bot users interact with many different bots and, therefore, are not able to spend significant effort on maintaining their configurations. Cogo and Hassan even suggest that there could be sharing platforms for distributing tailored configuration files for certain project characteristics.

Markusse et al. study the use of benchmarking bots in their paper titled ``Using Benchmarking Bots for Continuous Performance Assessment''. The authors show that bots are rarely used to continuously benchmark performance but that the situation is changing with the newly introduced GitHub Actions. Based on their findings, the authors encourage developers of performance-sensitive projects to consider adopting bot-based benchmarking. Specifically, adopting such bots can help with performance testing, the detection of performance regressions, and providing confidence to maintainers about complex changes. 

Golzadeh et al. make a call for better bot identification techniques in their paper entitled ``Recognizing bot activity in Collaborative Software Development''. The authors show that bots are among the most active accounts in open-source GitHub projects, yet they are rarely well identified. This large-spread presence of bots can impact certain analysis techniques that give credit based on activity. Hence, the authors argue that although current manual techniques tend to be the best option for maintainers, future work should examine the use of machine learning and artificial intelligence to detect bot activity.

\section{Future challenges}

The growth in popularity and contribution of bots is undeniable. The number of libraries, platforms, and reusable bots keeps mounting up. Nevertheless, to fully unleash the potential of bots in software engineering, we would like to draw attention to several technical and socioeconomic open challenges.

Regarding technical challenges, we need better systems to facilitate the coordination and collaboration of bots in the same project, as right now, each bot behaves in an independent way, and they can have conflicting actions. This challenge requires defining bot-specific coordination and integration policies.

Quality evaluation of bots is another key area. Mainly when bots include conversational capabilities, bot testing implies redefining many of the classical testing concepts as we need to test the behavioral part of the bot, the conversational component, and the combination of the two~\cite{RiccioJSHWT20, CabotBCDPR21}. 

Finally, security and privacy also pose relevant challenges. As we must be able to trust the bots we add to our projects, we need techniques that assure that bots will not perform malicious activities, leak data and request the bare minimum permissions.

Beyond technical aspects, we need to better understand users' perception of bots and how to optimize human-bot collaboration. Bots will need to get better communication and cognitive skills. For instance, when interacting with users, bots should be able to show empathy and react differently depending on the result of the sentiment analysis of the conversation they are having so far. Another example would be that the learn and mimic the specific idiosyncrasy of a project (including its vocabulary and natural language use) to increase their chances of being accepted. At the same time, bots could help in promoting social diversity in the project. As an example, they could identify and better support contributions from community minorities in the project. Finally, they should be able to explain their behavior to improve their trustworthiness.

The economic impact of bots in a project also deserves special attention. We do not have good economic models to evaluate the return-on-investment (ROI) of adopting a certain bot. If we could estimate the value of a bot for a project, it would be much easier to have rational discussions with project owners considering the cost-benefit analysis of integrating it. Even if some bots are released as open-source software, this doesn't mean there is no cost to adopt them. For example, developers often disregard the cost of learning how to use a bot properly.

So far, we have mostly discussed the impact of bots on software engineering. But as bots are software components themselves, bot development could and should benefit from well-grounded software engineering practices. What the best practices are for this specific type of software component remains to be seen. For example, it is still unclear how bots, especially collaborative and cognitive bots, will be tested. Bots are becoming smarter, and we know that the creation of smart software applications poses a specific set of additional challenges~\cite{Ozkaya20c}. We hope the community can benefit from this special issue's papers and keep working on innovating in this increasingly important field.
%this important support for software developers. 

\bibliographystyle{IEEEtran}
\bibliography{references}

\begin{wrapfigure}{l}{0.2\textwidth}
\includegraphics[width=0.2\textwidth]{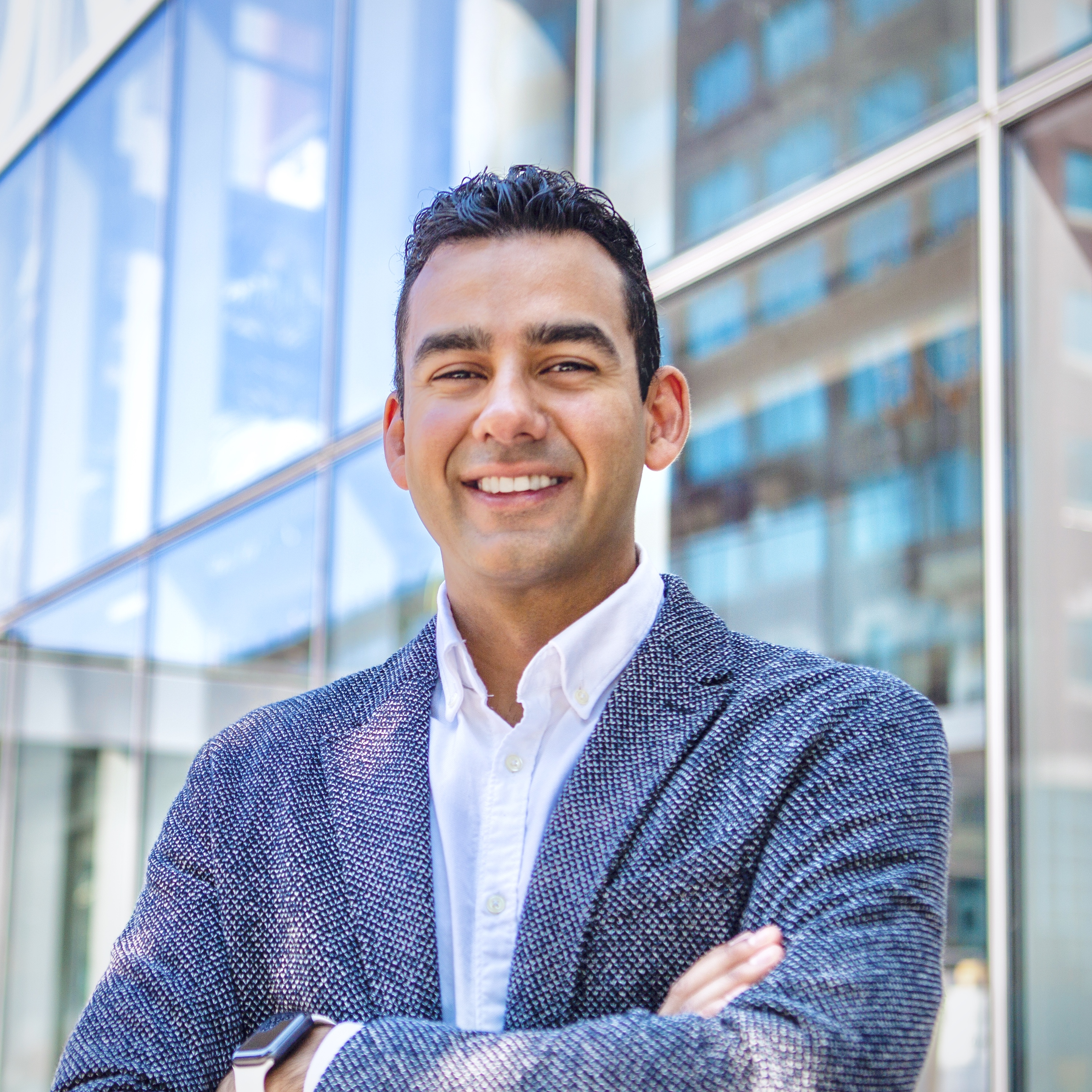}
\end{wrapfigure}
\begin{IEEEbiography}{Emad Shihab,}{\,}is a Full Professor and Research Chair at Concordia University where he leads the Data-driven Analysis of Software (DAS) lab. His research expertise is in Mining Software Repositories, Software Ecosystems, and Software Bots. He is a Senior Member of the IEEE. You can learn more at \url{https://das.encs.concordia.ca}.
\end{IEEEbiography}

\begin{wrapfigure}{l}{0.2\textwidth}
\includegraphics[width=0.2\textwidth]{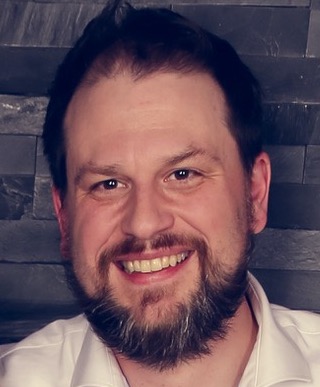}
\end{wrapfigure}
\begin{IEEEbiography}{Stefan Wagner}{\,} is a Full Professor of empirical software engineering and director of the Institute of Software Engineering at the University of Stuttgart. His research interests are human aspects, software quality, automotive software, AI-based systems, and empirical studies. He studied computer science in Augsburg and Edinburgh and received a doctoral degree from the Technical University of Munich. He is a senior member of IEEE and ACM. Contact him at \url{stefan.wagner@iste.uni-stuttgart.de}.\\\\
\end{IEEEbiography}

\begin{wrapfigure}{l}{0.2\textwidth}
\includegraphics[width=0.2\textwidth]{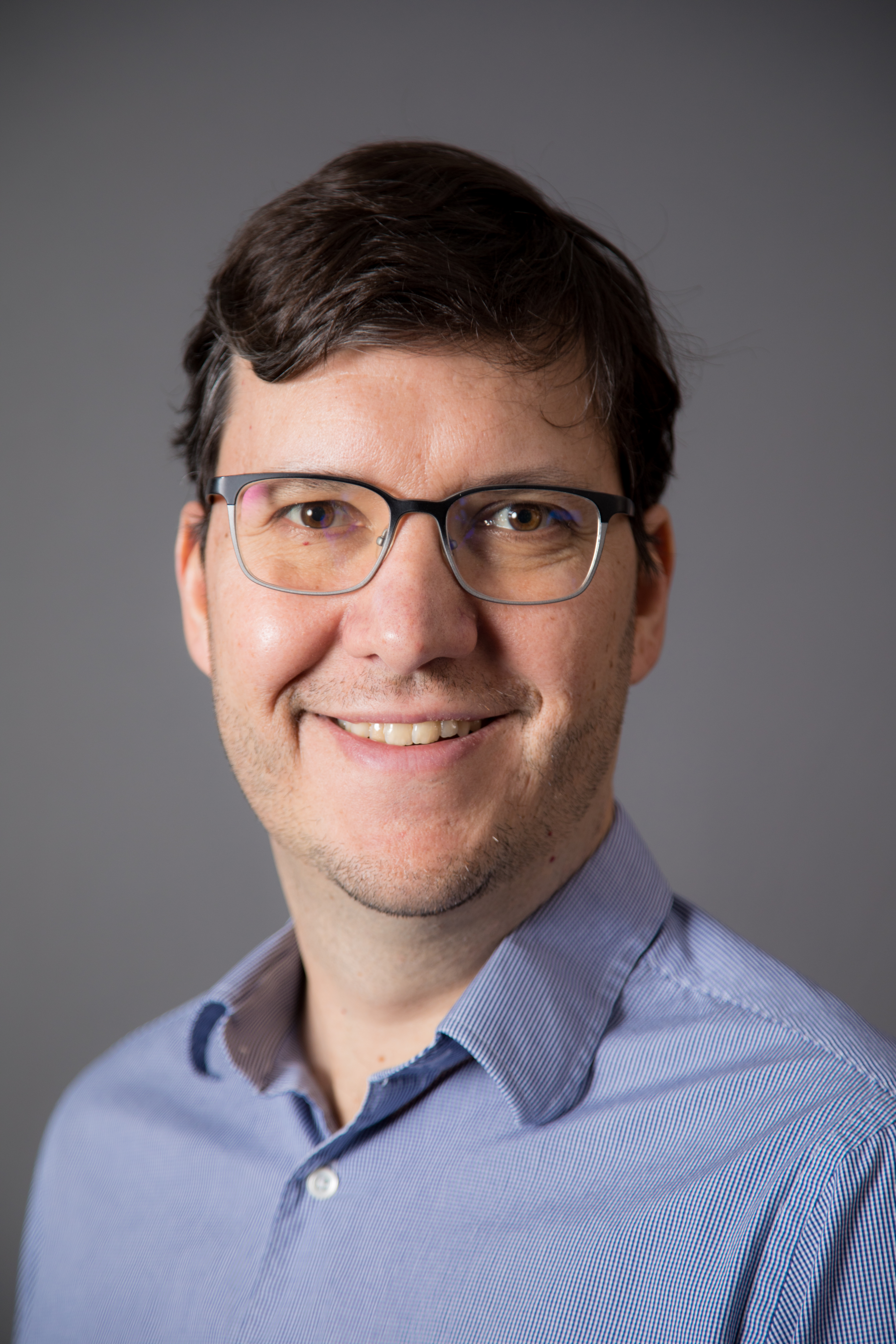}
\end{wrapfigure}
\begin{IEEEbiography}{Marco A. Gerosa}{\,} is a Full Professor at the Northern Arizona University, USA, and a Ph.D. advisor at the University of São Paulo, Brazil. He researches Software Engineering and CSCW. Recent projects include the development of tools and strategies to support newcomers' onboarding to open source software communities and the design of bots and chatbots. He published more than 200 papers and serves on the program committee (PC) of top-tier conferences, such as FSE, MSR, and CSCW. For more information, visit http://www.marcoagerosa.com
\end{IEEEbiography}

\begin{wrapfigure}{l}{0.2\textwidth}
\includegraphics[width=0.2\textwidth]{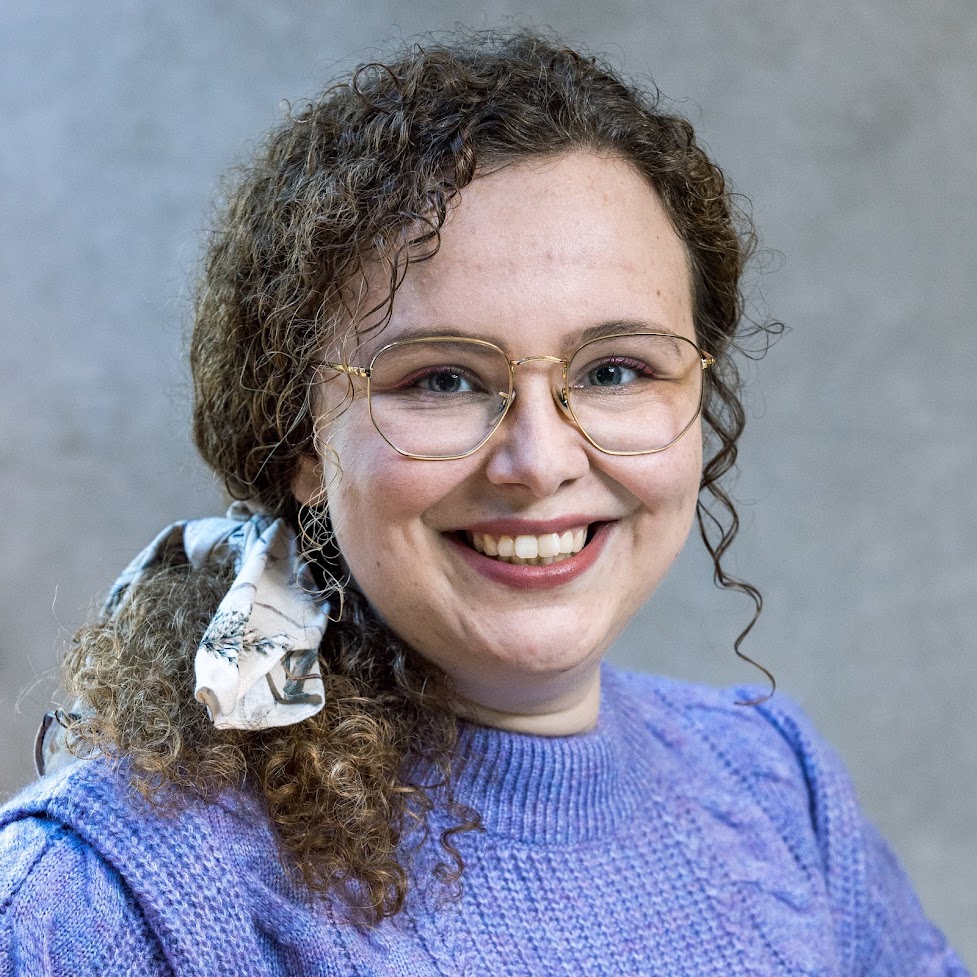}
\end{wrapfigure}
\begin{IEEEbiography}{Mairieli Wessel}{\,}is an Assistant Professor at the Radboud University, the Netherlands. She obtained her Ph.D. in Computer Science from the University of São Paulo, Brazil. Her main research interest is in software engineering (SE) and computer-supported cooperative work (CSCW), focused on software bots and open-source development. Her research goal is to design intelligent support for developers by leveraging bots’ capabilities. Contact her at \url{mairieli.wessel@ru.nl}.
\end{IEEEbiography}

\begin{wrapfigure}{l}{0.2\textwidth}
\includegraphics[width=0.2\textwidth]{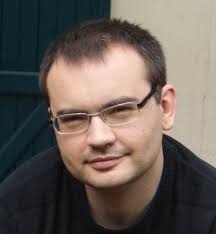}
\end{wrapfigure}
\begin{IEEEbiography}{Jordi Cabot}{\,} is an ICREA Research Professor at the Open University of Catalonia, Barcelona, Spain, E08035,  where he leads the Software and Systems Modeling Lab. Contact him at \url{jordi.cabot@icrea.cat} or \url{https://jordicabot.com}.
\end{IEEEbiography}

\end{document}